\documentclass[epj,numbook]{svjour}
\usepackage{graphics,epsfig,amsmath,amssymb}
\newcommand{\st}{{\bf S}_3(\pi)}
\begin{document}
\title{The 1D spin-$1\over2$ AF-Heisenberg model in a staggered field}
\authorrunning{Fledderjohann}
\dedication{Dedicated to J. Zittartz on the occasion of his 60th birthday.}
\author{A. Fledderjohann, M. Karbach and K.-H. M\"utter}
\institute{Physics Department, University of Wuppertal, D-42097 Wuppertal, Germany}
\date{\today}
\abstract{
  We investigate the scaling properties of the excitation energies and
  transition amplitudes of the one-dimensional spin-$1\over 2$ antiferromagnetic
  Heisenberg model exposed to an external perturbation. Two types of
  perturbations are discussed in detail: a staggered field and a dimerized field.
}
\maketitle
%
\section{Introduction}
\label{sec:Introduction}
%
The one-dimensional spin-$1\over2$ antiferromagnetic Heisenberg model with nearest
neighbour couplings and periodic boundary conditions (${\bf S}_{N+1} = {\bf
  S}_1$):
\begin{equation}
  \label{eq:H}
  {\bf H}\equiv 2\sum_{l=1}^N {\bf S}_l \cdot {\bf S}_{l+1} 
\end{equation} 
has been studied intensively with analytic and numerical methods.  Eigenvalues
as well as transition matrix elements for the spin operator have been calculated
with the Bethe ansatz and quantum group symmetries \cite{JM95}. These
calculations allow to exploit part of the dynamical properties of the model, the
two-spinon contributions, at $T=0$ \cite{KMB+97}. The dynamics of the model in
the presence of an external magnetic field, which is periodic in space
\begin{equation}
  \label{eq:Hh}
  {\bf H}(h) \equiv {\bf H}+ h{\bf S}_3(q)
\end{equation}
with 
\begin{equation}\label{eq:spi}
  {\bf S}_3(q)\equiv\sum_{l=1}^{N}e^{ilq}S_l^{3},
\end{equation}
has been studied so far mainly for $p=0$
\cite{FGM+96,MTBB81,JF86,KMS95,LR96,GKP+80,IS77}. Since the total spin ${\bf
  S}_3(0)$ commutes with ${\bf H}$, the eigenvectors of ${\bf H}(h)$ and ${\bf
  H}$ are the same, the eigenvalues change in a trivial manner.  For this reason
the magnetic properties of the model~(\ref{eq:H}) in a constant field -- as
there are magnetization curves, susceptibilities, static and dynamic structure
factors -- were accessible by means of the Bethe ansatz as well. The case of a
staggered field, i.e. Eq.~(\ref{eq:Hh}) for $p=\pi$, in a one-dimensional
Heisenberg model has attracted considerable interest as an effective model for
coupled spin chains. Treating the interchain coupling in the mean field
approach, reduces the system to a one-dimensional Heisenberg model in a
staggered field \cite{Schu96}.

It is the purpose of this paper to study excitation energies
\begin{equation}\label{eq:wmnNh}
  \omega_{mn}(N,h) \equiv E_m(N,h)-E_n(N,h),
\end{equation}
and transition amplitudes 
\begin{equation}\label{eq:smn}
  T_{mn}(N,h) \equiv \langle \Psi_m(h) | \st | \Psi_n(h) \rangle,
\end{equation}
for the operator $\st$ in the one-dimensional antiferromagnetic Heisenberg model
in a staggered field of strength $h$.

In section~\ref{sec:Evol-equat-excit} we derive a system of differential
equations, which describes the evolution in $h$ for $\omega_{mn}$ and $T_{mn}$.
Section \ref{sec:finite-size-scaling} is devoted to a finite-size scaling analysis
of the ground state excitations $\omega_{m0}$ and $T_{m0}$ in the combined limit
$h\to 0$ and $N\to\infty$. It is shown in section~\ref{sec:scaling-limit}, that the
critical exponents $\sigma$ and $\epsilon$ in the scaling ansatz and the scaling
variable as well can be computed by means of the evolution equations.  For small
values of the scaling variable the scaling function is also determined. In
section~\ref{sec:Heisenberg-dimer-model} we analyse in the same way the dimer
operator.
%
\section{Evolution equation for excitation energies and transition amplitudes}
\label{sec:Evol-equat-excit}
%
Starting from the eigenvalue equation of the Hamiltonian (\ref{eq:Hh})
\begin{eqnarray}
  \label{eq:HPsi}
  {\bf H}(h)|\Psi_n(h)\rangle &=& E_n(N,h)|\Psi_n(h)\rangle,
\end{eqnarray}
it is straightforward to derive  the evolution equations for the
energy eigenvalues and eigenstates:
\begin{eqnarray}\label{eq:diff-eq-a}
  \frac{d E_n(N,h)}{d h} &=&  T_{nn}(N,h), \\
  \frac{d}{dh}|\Psi_n(h)\rangle &=& 
   -\sum_{m\neq n}  \frac{T_{mn}(N,h)}{\omega_{mn}(N,h)} | \Psi_m(h) \rangle, 
  \label{eq:diff-eq-b}
\end{eqnarray}
where we have used the definitions~(\ref{eq:wmnNh}) and (\ref{eq:smn}).

The evolution is governed by the transition matrix elements of the operator $
\st$ [Eq. (\ref{eq:spi})]. The operator ${\bf S}_3(\pi)$ is real hermitian and
therefore the matrix (\ref{eq:smn}) is symmetric $T_{mn}=T_{nm}$.

Remarkably enough, there exists a closed system of differential equations, which
describes the $h$-dependence of the energy eigenvalues $E_{n}=E_n(N,h)$ and the
transition matrix elements $T_{mn}=T_{mn}(N,h)$:
\begin{eqnarray}
  \frac{d^{2} E_n}{dh^{2}} &=&  
   -2\sum_{l\neq n}\frac{|T_{ln}|^{2}}{\omega_{ln}}, \label{eq:d2En}\\
  \frac{d T_{mn}}{dh} &=&  
   \!-\!\sum_{l\neq m,n}\left[
    \frac{T_{ml}T_{ln}}{\omega_{lm}} + \frac{T_{ml}T_{ln}}{\omega_{ln}}
  \right]
  -\frac{T_{mn}}{\omega_{nm}}\frac{d\omega_{nm}}{dh}. \label{eq:dSmn}
 \nonumber \\
\end{eqnarray}
The Hamiltonian (\ref{eq:Hh}) behaves under translations ${\bf T}$ of one
lattice spacing as:
\begin{equation}
  \label{eq:THT}
  {\bf T} {\bf H}(h) {\bf T}^{\dag} = {\bf H}(-h).
\end{equation}
Since ${\bf T}$ is unitary, the eigenvalues 
\begin{equation}\label{eq:eh}
  E_n(N,h) = E_n(N,-h)
\end{equation}
are symmetric, whereas the eigenstates $|\Psi(h)\rangle$,
\begin{equation}\label{eq:psih}
  |\Psi(-h)\rangle = {\bf T} |\Psi(h)\rangle,
\end{equation}
are no longer momentum eigenstates, but linear combinations of two momentum
eigenstates with momenta $p=0$ and $p=\pi$, respectively. Of course, in the
limit $h\to 0$ translation invariance is recovered and $p$ is again a good
quantum number. We use the following notation for the energy eigenstates:
\begin{equation}\label{eq:Ppsi}
  {\bf T}|\Psi_n(0)\rangle = \pm |\Psi_n(0)\rangle, 
  \quad n=0,1,2,\ldots,
\end{equation}
which means these states carry momentum $p=0$ or $p=\pi$, respectively. The
ordering of the energy eigenstates is chosen in such a way that the ground state
$|\Psi_0(0)\rangle$ belongs to the ordered sequence of eigenstates with
$n=0,2,4,\ldots$. All eigenstates in this series have the same momentum. The
series of eigenstates with $n=1,3,5,\ldots$ carries the opposite sign. It starts
with the first excitation $|\Psi_1(0)\rangle$ and is independently ordered.

\begin{figure}[ht]
\centerline{\epsfig{file=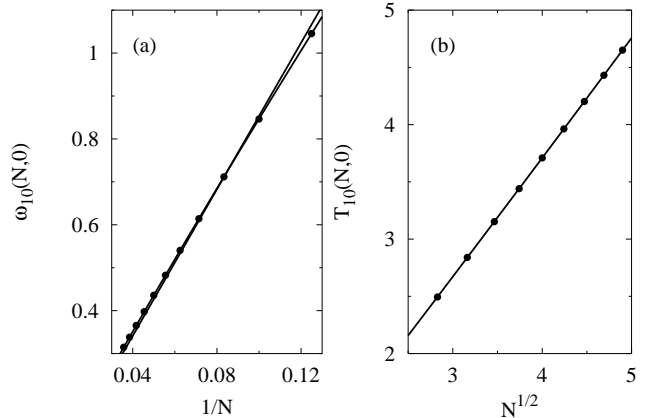,width=6.0cm,angle=-90}}
\caption{Finite-size dependence of the ground state energy (a) and the
  corresponding transition amplitude (b) of Hamiltonian~(\ref{eq:Hh}), for
  system sizes $N=8,10,\ldots,24\, (\bullet).$ The solid lines represent fits
  to the asymptotic behaviour~(\ref{eq:Em-E0-h0}) and~(\ref{eq:Sm0-h0}).}
\label{fig:s01-h0}
\end{figure}

Since the operator $\st$ changes the momentum of the state by $\pi$, we get
immediately from momentum conservation:
\begin{equation}\label{eq:Smn-h0}
  T_{mn}(N,0)=0 \quad \text{for} \quad (-1)^{m+n} = 1.
\end{equation}
The transition matrix elements $T_{m0}(N,0), m=1,3,5,\ldots$ from the ground
state $|\Psi_0(0)\rangle$ to the excited states $|\Psi_m(0)\rangle$ appear in
the definition of the dynamic structure factor
\begin{eqnarray}\label{eq:dsf}
  S(N,\omega,p=\pi,h=0) &=& \nonumber \\ && \hspace{-3cm} 
  \frac{1}{N}\sum_{m}\delta[\omega-\omega_{m0}(N,0)] 
  |T_{m0}(N,0)|^{2},
\end{eqnarray}
which is known to develop an infrared singularity in the thermodynamic
limit \cite{KMB+97}:
\begin{equation}\label{eq:Swpi-wto0}
  S(\infty,\omega,\pi,0) \stackrel{\omega\to 0}{\longrightarrow} 
  A\frac{\sqrt{\ln \omega}}{\omega}.
\end{equation}
The gap between the ground state and the excited states vanishes in the
thermodynamic limit:
\begin{equation}\label{eq:Em-E0-h0}
  \omega_{m0}(N,0) \stackrel{N\to\infty}{\longrightarrow} \frac{a_{m0}}{N}.
\end{equation}
The coefficients $a_{m0}$ can be computed in principle by the methods of
conformal invariance and finite-size scaling \cite{ABB87,WE87,ABB88}. The next
order terms entering in Eq.~(\ref{eq:Em-E0-h0}) are logarithmic corrections. For
this reason it becomes difficult to fix these coefficients from our system sizes
up to $N=24$. The difficulty of this task is extensively discussed in references
\cite{WE87,HBB88,ON92,NO93,Nomu93,KM95}. In figure~\ref{fig:s01-h0}(a) a fit
without logarithmic corrections ($a_{10}=8.53$) and with logarithmic corrections
($a_{10}=9.74$) is shown. The exact value is $a_{10}=\pi^{2}=9.86\ldots$
\cite{ABB88}.

Neglecting the logarithmic corrections in the large-$N$ behaviour, we expect for
the transition amplitudes: 
\begin{equation}\label{eq:Sm0-h0}
  T_{m0}(N,0) \stackrel{N\to\infty}{\longrightarrow} b_{m0} N^{\kappa},
  \quad m=1,3,\ldots.
\end{equation}
The exponent $\kappa$ turns out to be $\kappa\simeq 1/2$, as can be seen from
the numerical data for $T_{10}(N,0)$ in figure~\ref{fig:s01-h0}(b). This is in
agreement with conformal field theory, from which in the present case follows
$\kappa=a_{10}/2\pi^2$ (See \cite{Nijs81,CH93} and references therein). We
consider Eqs.  (\ref{eq:Smn-h0}), (\ref{eq:Em-E0-h0}) and (\ref{eq:Sm0-h0}) as
initial conditions for the system of differential equations (\ref{eq:d2En}) and
(\ref{eq:dSmn}).
%
\section{The finite-size scaling analysis}\label{sec:finite-size-scaling}
%
In this section we present numerical evidence for the validity of a finite-size
scaling ansatz:
\begin{equation}\label{eq:fsa-e} 
  \frac{\omega_{m0}(N,h)}{\omega_{m0}(N,0)} = 1 + e_{m0}(x) ,
\end{equation}
\begin{equation}\label{eq:fsa-s}
\frac{T_{m0}(N,h)}{T_{m0}(N,0)} = 1 + f_{m0}(x), \quad
  m=1,3,\ldots,  
\end{equation}
in the combined  limit 
\begin{equation}
  \label{eq:tl}
  N \longrightarrow \infty, \quad
  h \longrightarrow 0, \quad \text{with fixed} \quad x \equiv Nh^{\epsilon}.  
\end{equation}
Note, standard finite-size scaling theory \cite{PF83,PF84} usually employ
$y=x^{1/\epsilon}$ as the scaling variable. For our purpose it is more
convenient to use $x$ instead.  

\begin{figure}[ht]
  \centerline{\epsfig{file=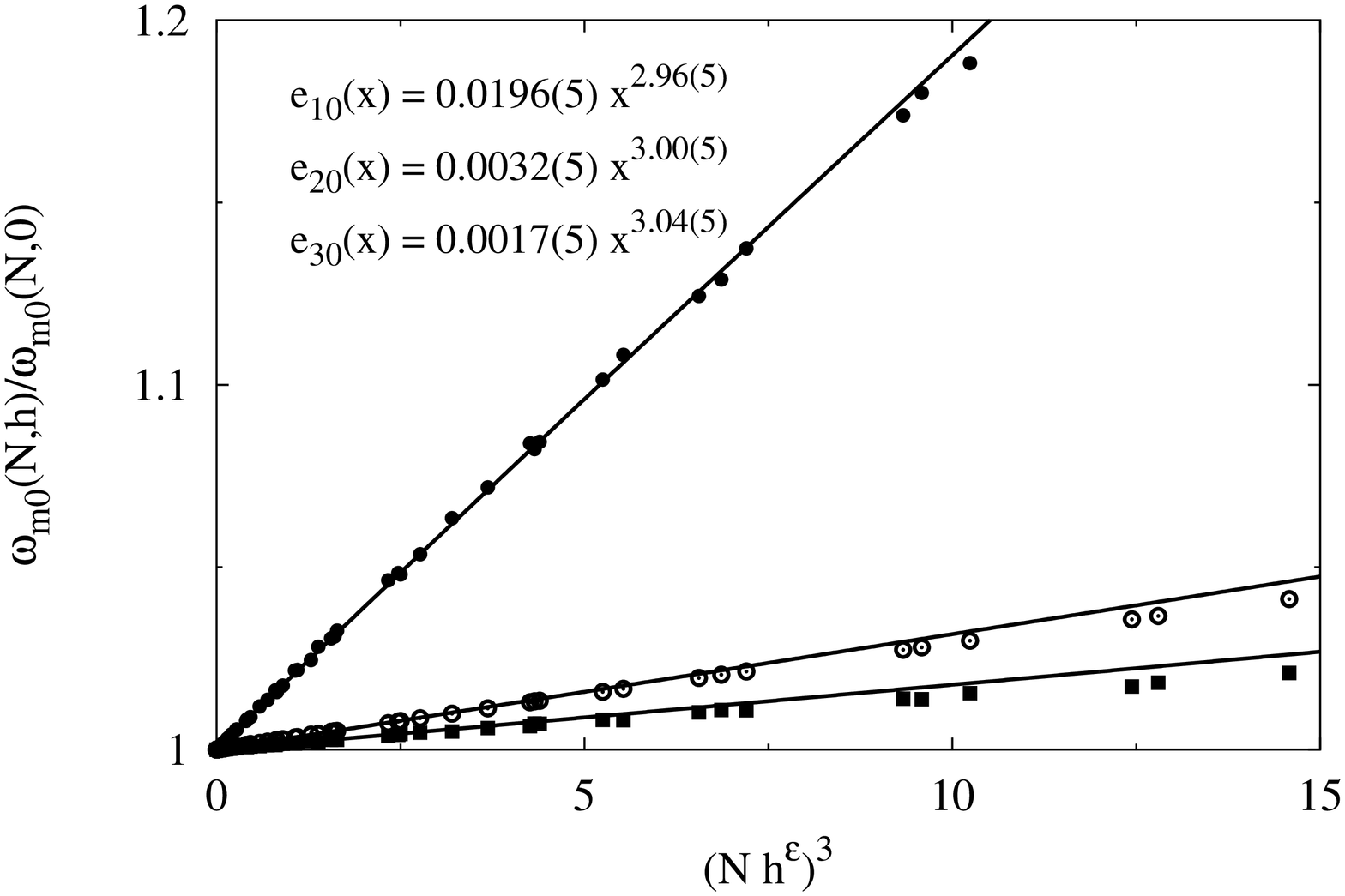,width=6cm,angle=-90}}
   \caption{The ratio~(\ref{eq:fsa-e}) for $m=1,2,3$ versus 
     $(Nh^{\epsilon})^{3}$ with $\epsilon=0.66$. The solid lines show the fits
     to the small $x$-behaviour of $e_{m0}(x)$ [cf. (\ref{eq:ex-fx-a})]. }
   \label{fig:fsa-e}
\end{figure}

\begin{figure}[ht]
  \centerline{\epsfig{file=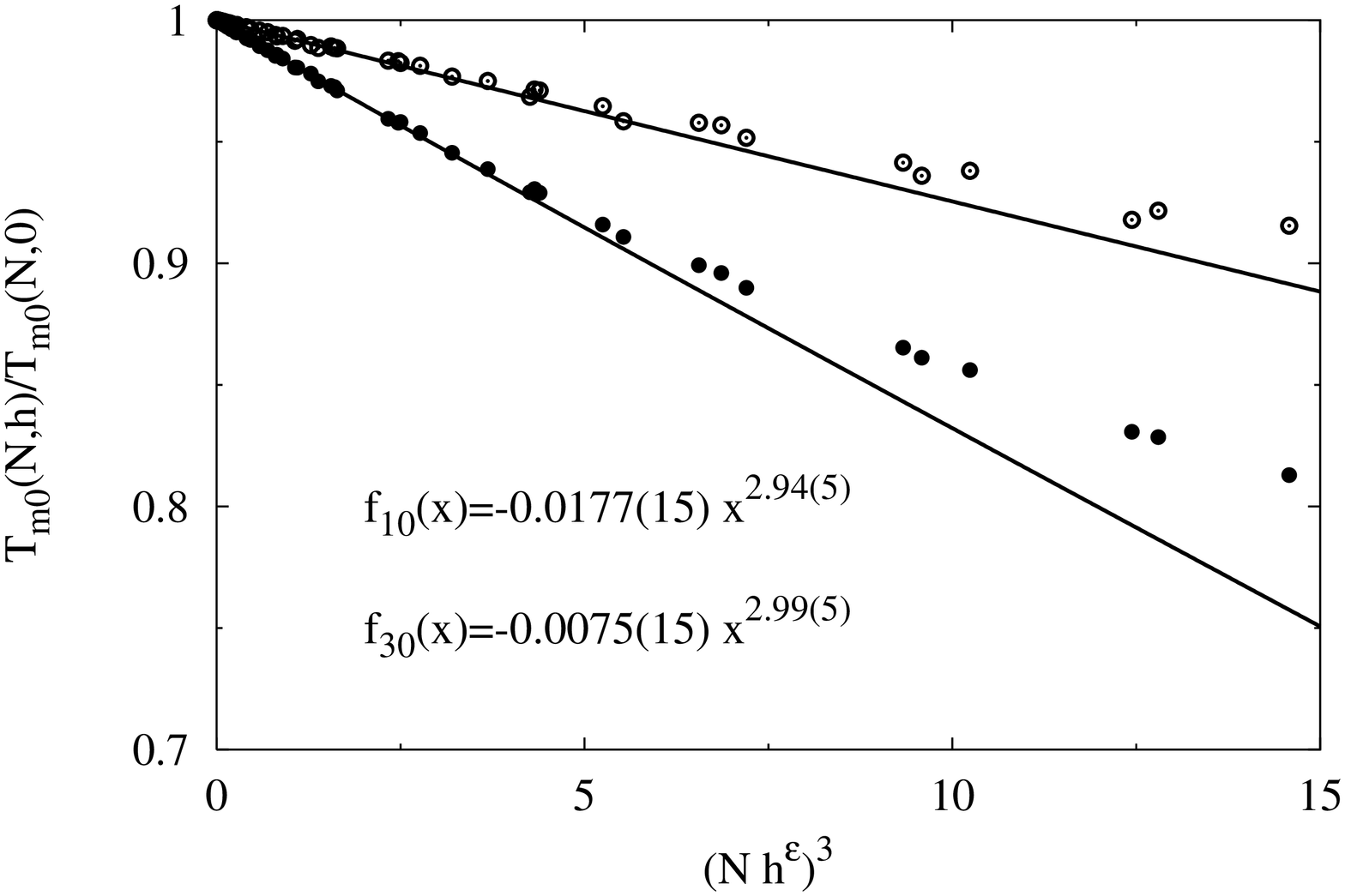,width=6cm,angle=-90}}
   \caption{The ratio~(\ref{eq:fsa-s}) for $m=1,3$ versus 
     $(Nh^{\epsilon})^{3}$ with $\epsilon=0.66$. The solid lines show the fits
     to the small $x$-behaviour of $f_{m0}(x)$ [cf. (\ref{eq:ex-fx-b})].}
   \label{fig:fsa-s}
\end{figure}

\begin{figure}[ht]
  \centerline{\hspace*{40mm}\epsfig{file=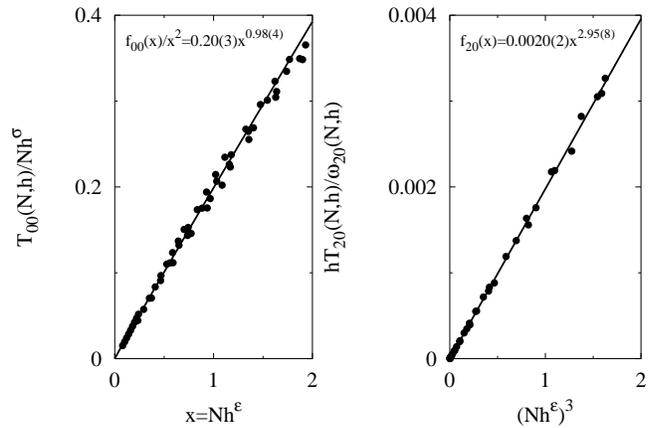,width=6cm,angle=-90}}
   \caption{The scaling behaviour of the transition amplitudes $T_{00}(N,h)$ and
     $T_{20}(N,h)$ for small $x$ [see Eqs.~(\ref{eq:S00-h}) and 
     (\ref{eq:S20-h})]. }
   \label{fig:s00-s20}
\end{figure}

In figures~\ref{fig:fsa-e} and~\ref{fig:fsa-s} we show our numerical results for
the ratios (\ref{eq:fsa-e}) and (\ref{eq:fsa-s}) on finite systems with
$N=8,10,...,24$. The best results are achieved for
\begin{equation}
  \label{eq:epsilon}
  \epsilon = 0.66(1).
\end{equation}
The small $x$-behaviour of the scaling functions is well parametrised by:
\begin{eqnarray}
  \label{eq:ex-fx-a} e_{m0}(x)&=&e_{m0}x^{\phi_e}, \\
  \label{eq:ex-fx-b} f_{m0}(x)&=&f_{m0}x^{\phi_f}, 
\end{eqnarray}
with
\begin{equation}
  \label{eq:phi}
  \phi_e = \phi_f = 3.00(4).
\end{equation}
The diagonal matrix elements of the staggered operator $\st$ vanish in the limit
$h\to 0$ [cf. Eq. (\ref{eq:Smn-h0})]. Therefore, a scaling ansatz of the type
(\ref{eq:fsa-s}) does not make sense. However, according to (\ref{eq:diff-eq-a})
the diagonal matrix elements of $\st$ can be identified with the first
derivative of the corresponding energy eigenvalue. Assuming, that these scale in
the same manner (\ref{eq:fsa-e}) as the excitation energies, we expect:
\begin{equation}
  \label{eq:S00-h}
  T_{00}(N,h) \rightarrow Nh^{\sigma}f_{00}(x)x^{-2},
\end{equation}
with $\sigma=1/3$. In figure~\ref{fig:s00-s20} we have plotted
$T_{00}(h)/Nh^{\sigma}$ versus the scaling variable $x=Nh^{\epsilon}$. The
scaling hypothesis works fairly well.

Moreover, the scaling function
\begin{equation}
  f_{00}(x)x^{-2} \stackrel{x\to 0}{\longrightarrow} f_{00}x,
\end{equation}
shows for small $x$-values a power behaviour, which is in agreement with
(\ref{eq:ex-fx-b}) and (\ref{eq:phi}). Finally, let us turn to the transition
amplitude $T_{20}(h)$, which also vanishes in the limit $h\to 0$. In
figure~\ref{fig:s00-s20} we have plotted 
\begin{equation}\label{eq:S20-h}
  h\frac{T_{20}(N,h)}{\omega_{20}(N,h)} = f_{20}(x),
\end{equation}
versus the scaling variable $x^{3}$. Here we have assumed that $T_{20}$ and
$\omega_{20}/h$ scale in the same manner. This is indeed the case as can be seen
in figure~\ref{fig:s00-s20}.  Moreover, in the small $x$ region the scaling
function behaves as it is expected from (\ref{eq:ex-fx-b}) and (\ref{eq:phi}).

In summary we observe that the excitation energies -- with respect to the ground
state -- scale with $h^{\epsilon}\; (\epsilon\simeq 2/3)$:
\begin{equation}
  \label{eq:e-gap}
  \omega_{m0}(N,h) = h^{\epsilon} \Omega_{m0}(x).
\end{equation}
As before, we have neglected logarithmic corrections due to marginal operators.
These corrections were deduced by means of conformal field theory \cite{AGSZ89}.
The transition amplitudes scale with $h^{\sigma}N\; (\sigma\simeq 1/3)$:
\begin{equation}
  \label{eq:s-gap}
  T_{m0}(N,h) = N h^{\sigma} \Theta_{m0}(x).
\end{equation}
The small $x$-behaviour of the scaling functions $\Omega_{m0}(x)$ and
$\Theta_{m0}(x)$ is fixed by the initial conditions (\ref{eq:Em-E0-h0}) and
~(\ref{eq:Sm0-h0}):
\begin{equation}
  \label{eq:Theta_n0}
  \Omega_{m0}(x) = a_{m0} x^{-1}[1+ e_{m0}(x)],\quad 
  m=0,1,2,\ldots 
\end{equation}
According to (\ref{eq:Smn-h0}) and (\ref{eq:Em-E0-h0}) the small $x$-behaviour of
different $\Theta_{m0}(x)$ is different for $m$ even and $m$ odd, respectively:
\begin{equation} \label{eq:Sigma_n0}
  \Theta_{m0}(x) = \left\{
  \begin{array}{ll}
    a_{m0}x^{-2}f_{m0}(x)     \; &: (-1)^{m}=1 \\
    b_{m0}x^{-1/2}[1+f_{m0}(x)]\; &: (-1)^{m}=-1.
  \end{array}\right.
\end{equation}
Again $b_{m0}$ is given by the initial condition (\ref{eq:Sm0-h0}). The large
$x$-behaviour of the scaling function for the first excitation is of special
interest. If
\begin{equation}
  \label{eq:5}
  \lim_{x\to\infty}\Omega_{10}(x)=\Omega_{10}^{\infty}
\end{equation}
is finite, Eq.~(\ref{eq:e-gap}) tells us that there is a gap in the
thermodynamic limit, which opens with $h^{\epsilon}, \epsilon\simeq 2/3$.  We
have analysed our finite system data with the BST \cite{BS64} algorithm and
found:
\begin{equation}
  \label{eq:h-gap-x-infinty}
  \Omega_{10}^{\infty}=4.
\end{equation}
A similar statement can be derived from Eq.~(\ref{eq:s-gap}) for the staggered
magnetization, which can be expressed in terms of the transition matrix elements
$T_{m0}(N,h)$:
\begin{equation}
  \label{eq:stag-mag}
  m^{\dag}(h) = \sum_{m}|T_{m0}(N,h)|^{2}.
\end{equation}
%
\section{The evolution equation in the scaling limit}
\label{sec:scaling-limit}
%
The critical exponents $\sigma=1/3$ and $\epsilon=2/3$ in Eqs.~(\ref{eq:e-gap})
and (\ref{eq:s-gap}) as well as the small $x$-behaviour $\phi=3$ of the scaling
functions shown in figures~\ref{fig:fsa-e},~\ref{fig:fsa-s} and \ref{fig:s00-s20}
will now be derived from the evolution equations~(\ref{eq:diff-eq-a}) and
(\ref{eq:diff-eq-b}) in the scaling limit (\ref{eq:tl}). We assume that a scaling
ansatz of the type (\ref{eq:e-gap}) and (\ref{eq:s-gap})
\begin{eqnarray}
  \label{eq:e-s-gap-a}\omega_{mn}(N,h) &=& h^{\epsilon} \Omega_{mn}(x), \\
  \label{eq:e-s-gap-b}     T_{mn}(N,h) &=& N h^{\sigma} \Theta_{mn}(x), 
\end{eqnarray}
with the scaling variable
\begin{equation}
  x=Nh^{\epsilon},
\end{equation}
holds for all the transition energies and amplitudes. Furthermore, let us assume
that the small $x$-behaviour of the scaling functions:
\begin{equation}
  \label{eq:Omega-x-to-0}
  \Omega_{mn}(x) \stackrel{x\to 0}{\longrightarrow} a_{mn} x^{-1}[1+e_{mn}(x)],
\end{equation}
\begin{equation}
  \label{eq:Xi-x-to-0}
  \Theta_{mn}(x) \stackrel{x\to 0}{\longrightarrow} 
  \left\{\begin{array}{ll}
      a_{mn}x^{-2} f_{mn}(x) &:  (-1)^{n+m}=1 \\
      b_{mn}x^{-\sigma/\epsilon} [1+f_{mn}(x)] &:  (-1)^{n+m}=-1
      \end{array}       
  \right.,
\end{equation}
follows from the initial conditions:
\begin{equation}
 \lim_{N\to\infty} N\omega_{mn}(N,0) = a_{mn},
\end{equation}
and
\begin{equation}\label{eq:Smnh0-N-to-infty}
  T_{mn}(N,0) \stackrel{N\to\infty}{\longrightarrow}b_{mn}N^{\kappa}
\end{equation}
where $ b_{mn}=0$ for $(-1)^{m+n}=+1 $.  The exponents $\kappa,\sigma$ and
$\epsilon$ are then related:
\begin{equation}
  \label{eq:kappa-sigma-epsilon}
  \kappa= 1 -\frac{\sigma}{\epsilon}.
\end{equation}
Now we discuss the evolution equations (\ref{eq:d2En}) and (\ref{eq:dSmn}) in the
scaling limit~(\ref{eq:tl}), in conjunction with the scaling ansatz
[(\ref{eq:e-s-gap-a}) and (\ref{eq:e-s-gap-b})].  It is convenient to inspect the
derivative
\begin{equation}
  \label{eq:d2Emn}
   D_1 = N^{-2}h^{-3\epsilon+2} \frac{d^{2}}{dh^{2}} \omega_{mn}(N,h).
  \end{equation}
  Then the left- and right-hand sides of (\ref{eq:d2En}) acquire the following
  form:
\begin{eqnarray}\label{eq:d2Emn-lhs}
  D_1 &=&  \epsilon(\epsilon\!-\!1)x^{-2}\Omega_{mn}(x) \nonumber \\
   && +\epsilon(3\epsilon\!-\!1)x^{-1}
    \frac{d\Omega_{mn}(x)}{dx}+ \epsilon^{2}\frac{d^{2}\Omega_{mn}(x)}{dx^{2}},
  \\ \label{eq:d2Emn-rhs}
  D_1 &=& 2h^{2\sigma-4\epsilon+2} \nonumber \\ && \times 
  \left[
    2\frac{\Theta_{mn}^{2}(x)}{\Omega_{mn}(x)} 
    -\sum_{l\neq m,n}\frac{\Theta_{lm}^{2}(x)}{\Omega_{lm}(x)}
    +\sum_{l\neq m,n}\frac{\Theta_{ln}^{2}(x)}{\Omega_{ln}(x)}
  \right]. \nonumber \\
  \end{eqnarray}
A similar calculation for  the derivative:
\begin{equation}
  \label{eq:dlnS_mn-dh}
  D_2 = h \frac{d}{dh} \ln\left[T_{mn}(N,h)\omega_{mn}(N,h)\right]
\end{equation}
yields for the left and right-hand sides of Eq.~(\ref{eq:dSmn}):
\begin{eqnarray}\label{eq:dlnS_mn-dh-lhs}
  D_2 &=& (\sigma+\epsilon)
 + \epsilon x \frac{d}{dx} \ln[\Theta_{mn}(x)\Omega_{mn}(x)], 
 \\ \label{eq:dlnS_mn-dh-rhs}
  D_2 &=& -h^{\sigma-2\epsilon+1}x \nonumber \\ && \times
  \sum_{l\neq m,n}
  \frac{\Theta_{ml}(x)\Theta_{ln}(x)}{\Theta_{mn}(x)}
  \left[ \frac{1}{\Omega_{lm}(x)} +  \frac{1}{\Omega_{ln}(x)}  \right].
 \nonumber \\
\end{eqnarray}
The corresponding left-hand sides and right-hand sides in (\ref{eq:d2Emn}) and
(\ref{eq:dlnS_mn-dh}) scale in the same manner provided that 
\begin{equation}\label{eq:sigma-epsilon}
  \sigma = 2\epsilon-1.
\end{equation}
In this case (\ref{eq:e-s-gap-a}) and (\ref{eq:e-s-gap-b}) lead to a system of
differential equations for the scaling functions $\Omega_{mn}(x)$ and
$\Theta_{mn}(x)$.

Combining (\ref{eq:sigma-epsilon}) and (\ref{eq:kappa-sigma-epsilon}) we see
that the exponents
\begin{equation}
  \epsilon=\frac{1}{1+\kappa},\quad \sigma=\frac{1-\kappa}{1+\kappa},
\end{equation}
are fixed by the exponent $\kappa$ in the initial conditions
(\ref{eq:Smnh0-N-to-infty}).  

The small $x$-behaviour is governed by the initial conditions (\ref{eq:Sm0-h0})
and (\ref{eq:Em-E0-h0}). They tell us that the leading behaviour of the
right-hand side of (\ref{eq:d2Emn-rhs}) arise from excitations with
$(-1)^{l+m}=-1$ and $(-1)^{l+n}=-1$
\begin{equation}\label{eq:x-0-Xi-Omega}
  \lim_{x\to 0} \frac{\Theta^{2}_{lm}(x)}{\Omega_{lm}(x)} = 
  \frac{b_{lm}^{2}}{a_{lm}}x^{1-2\sigma/\epsilon}.
\end{equation}
Therefore, the terms in (\ref{eq:d2Emn-rhs}) should be proportional to
$x^{-3+2/\epsilon}$. Assuming that the small $x$-behaviour of functions
$e_{mn}(x)$ can be described by a power law ansatz:
\begin{equation}
  e_{mn}(x) = e_{mn}x^{\phi_e} + \ldots,
\end{equation}
it follows from (\ref{eq:d2Emn-lhs}) that
\begin{equation}
  D_1 = a_{mn}e_{mn}x^{\phi_e-3}\phi_e\epsilon(\phi_e\epsilon-1)
\end{equation}
is also proportional to $x^{-3+2/\epsilon}$ if
\begin{equation}
   \phi_e=\frac{2}{\epsilon}.
\end{equation}
Therefore, the first coefficient $e_{mn}$ in the scaling function $e_{mn}(x)$ is
completely fixed by the initial values $a_{ln},b_{lm}$:
\begin{eqnarray}\label{eq:e_mn}
  2e_{mn} &=& 2\left(1-(-1)^{m+n}\right)\frac{b_{mn}^{2}}{a_{mn}^{2}} 
  \nonumber \\ && \hspace{-1cm} -\frac{1}{a_{mn}} 
   \sum_{l\neq m,n} \left[
     \left(1\!-\!(-1)^{l+m}\right)\frac{b_{lm}^{2}}{a_{lm}} -
   \left(1\!-\!(-1)^{l+n}\right)\frac{b_{ln}^{2}}{a_{ln}}
   \right]. \nonumber \\
\end{eqnarray}
Next we study the small $x$-behaviour of (\ref{eq:dlnS_mn-dh}). We will discuss
the two cases $(-1)^{m+n}=1$ and $(-1)^{m+n}=-1$ separately.
\subsection{$(-1)^{m+n}=1$}\label{sec:scaling-limit-A}
The right-hand side of  (\ref{eq:dlnS_mn-dh-rhs}) is governed by the excitations
with $(-1)^{l+m}=-1$ and  $(-1)^{l+n}=-1$:
\begin{eqnarray}\label{eq:A-x-to-0-rhs}
  x\frac{\Theta_{ml}(x)\Theta_{ln}(x)}{\Theta_{mn}(x)}
  \left[\frac{1}{\Omega_{lm}(x)}+\frac{1}{\Omega_{ln}(x)}\right] \nonumber \\
  && \hspace{-5cm} \stackrel{x\to 0}{\longrightarrow} 
  \frac{b_{ml}b_{ln}}{a_{mn}}\left[\frac{1}{a_{lm}}+\frac{1}{a_{ln}}\right]
  \frac{x^{2/\epsilon}}{f_{mn}(x)},
\end{eqnarray}
For the left-hand side of (\ref{eq:dlnS_mn-dh-lhs}) we get:
\begin{eqnarray}\label{eq:A-x-to-0-lhs}
  3\epsilon-1+\epsilon x 
  \frac{d}{dx} \ln[\Theta_{mn}(x) \Omega_{mn}(x)] \nonumber && \\
  && \hspace{-5cm}  \stackrel{x\to 0}{\longrightarrow} 
    -1 + \epsilon x \frac{d}{dx} \ln f_{mn}(x).
\end{eqnarray}
Consistency of (\ref{eq:A-x-to-0-rhs}) and (\ref{eq:A-x-to-0-lhs}) is achieved if
\begin{equation}\label{eq:f_mn_x}
  f_{mn}(x) \stackrel{x\to 0}{\longrightarrow} f_{mn}x^{\phi_f},
\end{equation}
with $\phi_f=\phi_e=2/\epsilon$. The coefficient $f_{mn}$ in (\ref{eq:f_mn_x}) is
completely fixed by the initial conditions:
\begin{eqnarray}\label{eq:f_mn}
  f_{mn} &=& -\sum_{l\neq m,n} \frac{1-(-1)^{l+m}}{2} \frac{1-(-1)^{l+n}}{2}
  \nonumber \\ && \times 
  \frac{b_{ml}b_{ln}}{a_{mn}} \left(\frac{1}{a_{lm}}+\frac{1}{a_{ln}}\right).
\end{eqnarray}
\subsection{$(-1)^{m+n}=-1$}\label{sec:scaling-limit-B}
The contributions in (\ref{eq:dlnS_mn-dh-rhs}) vanish:
\begin{eqnarray}\label{eq:B-x-to-0-rhs}
  x\frac{\Theta_{ml}(x)\Theta_{ln}(x)}{\Theta_{mn}(x)}
  \left[\frac{1}{\Omega_{lm}(x)}+\frac{1}{\Omega_{ln}(x)}\right] \nonumber \\
  && \hspace{-5cm} \stackrel{x\to 0}{\longrightarrow} 
  \frac{1+(-1)^{m+l}}{2} \frac{b_{ln}a_{ml}}{b_{mn}} 
  \left[\frac{1}{a_{ln}}+\frac{1}{a_{lm}}\right]f_{ml}(x) \nonumber \\
  && \hspace{-4.5cm} +
  \frac{1+(-1)^{n+l}}{2} \frac{b_{ml}a_{ln}}{b_{mn}} 
  \left[\frac{1}{a_{ml}}+\frac{1}{a_{nl}}\right]f_{ln}(x),\nonumber \\
\end{eqnarray}
since the scaling function $f_{lm}$ vanishes for the excitations with
$(-1)^{l+m}=1$, according to equation (\ref{eq:f_mn_x}). The leading behaviour
of (\ref{eq:dlnS_mn-dh-lhs}):
\begin{eqnarray}\label{eq:B-x-to-0-lhs}
  3\epsilon-1+\epsilon x \frac{d}{dx} 
    \ln[\Theta_{mn}(x) \Omega_{mn}(x)] \nonumber && \\
  && \hspace{-5cm}  \stackrel{x\to 0}{\longrightarrow} 
    \epsilon \left(\frac{d}{dx}f_{mn}(x)+e_{mn}x^{2/\epsilon}\right)
\end{eqnarray}
is governed by the scaling functions $e_{mn}(x)$ and $f_{mn}(x)$ for the
excitations with $(-1)^{m+n}=-1$. Combining (\ref{eq:B-x-to-0-rhs}) and
(\ref{eq:B-x-to-0-lhs}) we obtain:
\begin{eqnarray}\label{eq:e_mn-f_mn}
  2 (e_{mn}+f_{mn}) && 
  \nonumber \\ && \hspace{-2cm} 
  =\sum_{l\neq m,n} \frac{1+(-1)^{l+m}}{2} f_{ml}
  \frac{b_{ln}a_{ml}}{b_{mn}} \left[\frac{1}{a_{ml}}+\frac{1}{a_{ln}}\right] 
  \nonumber \\ && \hspace{-2cm} 
  +\sum_{l\neq m,n} \frac{1+(-1)^{l+n}}{2} f_{ln}
  \frac{b_{ml}a_{ln}}{b_{mn}} \left[\frac{1}{a_{ml}}+\frac{1}{a_{ln}}\right] 
\end{eqnarray}
In summary we conclude:
\begin{enumerate}
\item For consistency, all the scaling functions:
  \begin{eqnarray}
    e_{mn}(x) &=& e_{mn}x^{\phi}+e_{mn}^{(1)}x^{\phi_1}+\ldots, \\
    f_{mn}(x) &=& f_{mn}x^{\phi}+f_{mn}^{(1)}x^{\phi_1}+\ldots, 
  \end{eqnarray}
  have the same small $x$-behaviour with exponent $\phi=2/\epsilon$.
 
\item Equations (\ref{eq:e_mn}), (\ref{eq:f_mn}) and (\ref{eq:e_mn-f_mn}) enable
  us to express the first coefficients $e_{mn},f_{mn}$ in terms of the initial
  values $a_{lm},b_{lm}$.
  
\item Having determined the lowest order in $x$ of the scaling functions, we can
  proceed and compute the next order $x^{\phi_1}$ from the differential
  equations (\ref{eq:d2Emn}) and (\ref{eq:dlnS_mn-dh}). The exponent $\phi_1$
  turns out to be $\phi_1=2\phi=4/\epsilon$ for all the scaling functions.
  Moreover, the next order coefficients $e_{mn}^{(1)},f_{mn}^{(1)}$ can be
  expressed again in terms of the initial values $a_{mn},b_{mn}$ and the zeroth
  order coefficients $e_{mn},f_{mn}$.
  
\item The power ansatz ansatz~(\ref{eq:Omega-x-to-0}) and~(\ref{eq:Xi-x-to-0})
  for the small $x$-behaviour of the scaling function is only consistent with a
  power ansatz~~(\ref{eq:Smnh0-N-to-infty}) for the finite-size corrections of
  the initial values. Logarithmic corrections of the latter will induce
  logarithmic corrections in the small $x$-behaviour of the scaling functions.

\end{enumerate}
Finally let us discuss equations (\ref{eq:e_mn}) and~(\ref{eq:e_mn-f_mn}) for
the lowest excitations $e_{10}$ and $f_{10}$ from the ground state. The
numerical results shown in figures~\ref{fig:fsa-e} and~\ref{fig:fsa-s} tell us
that the coefficients in the scaling functions for the higher excitations
$e_{03}$ and $f_{03}$ are considerably smaller. If we neglect in equations
(\ref{eq:e_mn}) and (\ref{eq:e_mn-f_mn}) all higher excitations we find
\begin{equation}
  \label{eq:e01=2b01h2/a01}
  e_{10} = 2 \frac{b_{10}^{2}}{a_{10}^{2}},
\end{equation}
and 
\begin{equation}
  \label{eq:e01+f01}
  e_{10}+f_{10} \simeq 0.
\end{equation}
A comparison of the slopes 
\begin{equation}
  \label{eq:e01f01-num}
  e_{10}=0.0196(8),\quad f_{10}=-0.0177(8),
\end{equation}
in figures~\ref{fig:fsa-e} and~\ref{fig:fsa-s} demonstrates that
Eq.~(\ref{eq:e01+f01}) is indeed well satisfied. The numerical values of
$a_{10}= 9.74(8)$, $b_{10}=0.95(3)$, estimated from the data of
figure~\ref{fig:s01-h0}, together with Eq.~(\ref{eq:Omega-x-to-0}) yield a value
$e_{10}=0.019$. This is in agreement with Eq.~(\ref{eq:e01f01-num}) and might be
seen as an indication that the neglect of higher excitations is justified.
%
\section{The Heisenberg model in a \\ dimerized field}
\label{sec:Heisenberg-dimer-model}
%
The general features developed for the 1D spin-1/2 Heisenberg model in a
staggered field can be applied to all Hamiltonians of the type
\begin{equation}
  \label{eq:H-delta}
  {\bf H}(\delta) \equiv {\bf H} + \delta {\bf H}_D
\end{equation}
where ${\bf H}$ is the unperturbed Hamiltonian (\ref{eq:H}) and ${\bf H}_D$ 
denotes a perturbation operator. As a further interesting example we want to
discuss here the dimer operator 
\begin{equation}
  \label{eq:dimer-op}
  {\bf H}_D = 2\sum_{n=1}^{N} (-1)^{n} {\bf S}_n \cdot {\bf S}_{n+1} .
\end{equation}
Hamiltonians of type (\ref{eq:H-delta}) with ${\bf H}_D$ as~(\ref{eq:dimer-op})
have been discussed in the framework of (organic) spin-Peierls materials
\cite{JBH+76}, where the interaction between linear antiferromagnetic chains and
the three-dimensional phonon system causes a dimerization of the lattice
\cite{Pytt74,BBK78,CF79}. In case of the more recently discovered inorganic
spin-Peierls material $\rm CuGeO_3$, reference~\cite{HTU93}, it turned out that
in addition a next nearest neighbour coupling in the spin chain has to be
considered \cite{FKL+98}. We will not discuss the influence of such a term in
this paper.

The evolution equation for the energy eigenvalues $E_n(\delta) $ and the
transition amplitudes:
\begin{equation}
  \label{eq:Smn-delta}
  T_{mn}(N,\delta) = \langle \Psi_m(\delta) | {\bf H}_D | \Psi_n(\delta) \rangle
\end{equation}
are obtained from (\ref{eq:d2En}) and (\ref{eq:dSmn}) by a simple
substitution: $h\mapsto \delta,\;E_n(N,h)\mapsto E_n(N,\delta)$, and
$T_{mn}(N,h) \mapsto T_{mn}(N,\delta)$. Note however, that the dimer
operator~(\ref{eq:dimer-op}) conserves the total spin in contrast to the
staggered spin operator. Therefore, the excitations from the ground state are
singlet states as well.

\begin{figure}[ht]
  \centerline{\hspace*{40mm}
       \epsfig{file=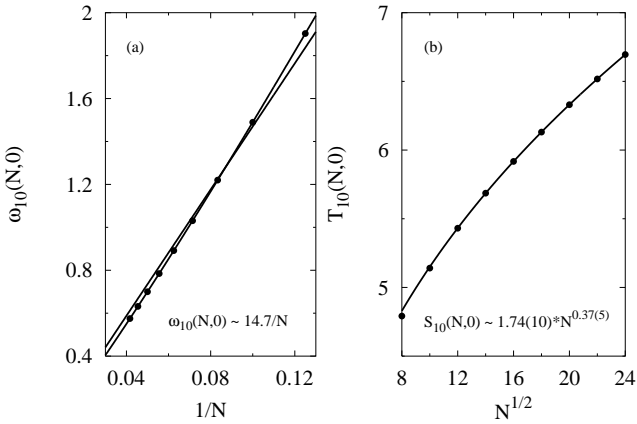,width=6cm,angle=-90}}
   \caption{Finite-size dependence of the ground state energy (a) and the
  corresponding transition amplitude (b) of Hamiltonian~(\ref{eq:H-delta}), for
  system sizes $N=8,10,\ldots,24\, (\bullet).$ The solid lines represent fits to
  the asymptotic behaviour~(\ref{eq:Em-E0-h0}) and~(\ref{eq:Sm0-h0}).}
   \label{fig:S01-delta0}
\end{figure}

For the solution of the evolution equation we need the initial values for the
excitation energies:
\begin{equation}
  \label{eq:Emn-delta-0}
  \lim_{N\to\infty}N\omega_{mn}(N,\delta=0)= a_{mn},
\end{equation}
and for the transition amplitudes:
\begin{eqnarray}  
  T_{mn}(N,0) &=& 0 \quad\mbox{for}\quad (-1)^{m+n}=1 ,
   \label{eq:Smn-delta-0+1} \\
  T_{mn}(N,0) &=& b_{mn} N^{\kappa}\quad\mbox{for}\quad (-1)^{m+n}=-1.
  \label{eq:Smn-delta-0-1} 
\end{eqnarray}
Equation (\ref{eq:Smn-delta-0+1}) results from momentum conservation at
$\delta=0$ and the fact that the operator (\ref{eq:dimer-op}) changes the
momentum by $\pi$. Equation (\ref{eq:Smn-delta-0-1}) is an ansatz for the
finite-size dependence of the transition amplitudes. Numerical data are shown
for $T_{10}(N,0)$ in figure~\ref{fig:S01-delta0}. They suggest an exponent
$\kappa=0.37(5)$, which deviates from the standard values $\kappa=1/2$,
predicted by conformal field theory \cite{Nijs81,ABB88}. The discrepancy is due
to strong logarithmic corrections, which are disregarded in the
fit~(\ref{eq:Smn-delta-0-1}). This leads to a shift of the exponent $\kappa$. We
expect that the standard value $\kappa=1/2$ will show up with increasing system
size.

\begin{figure}[ht]
  \centerline{\hspace*{40mm}\epsfig{file=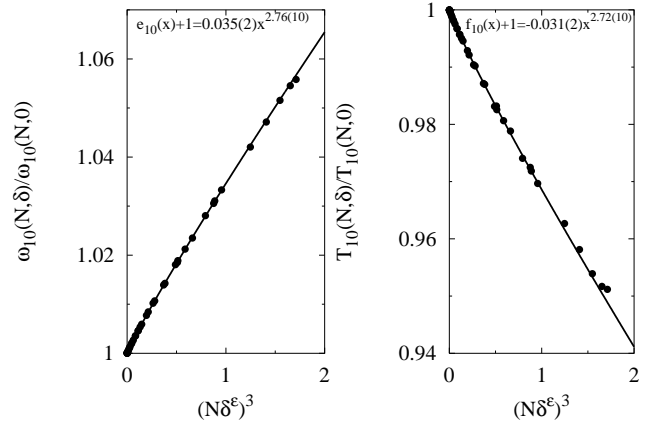,width=6cm,angle=-90}}
   \caption{The ratios~(\ref{eq:fsa-e01}) and (\ref{eq:fsa-s01}) for 
            $e_{10}(x)$ and $f_{10}(x)$ versus $(N\delta^{\epsilon})^{3}$ with 
            $\epsilon=0.73$. The solid lines show the fits to the small 
            $x$-behaviour.}
   \label{fig:e01s01-delta}
\end{figure}

Next we study the finite-size scaling ansatz of the type (\ref{eq:fsa-e}) and
(\ref{eq:fsa-s}):
\begin{eqnarray}
  \frac{\omega_{10}(N,\delta)}{\omega_{10}(N,0)} &=& 1 + e_{10}(x) 
  \label{eq:fsa-e01} \\
\frac{T_{10}(N,\delta)}{T_{10}(N,0)} &=& 1 + f_{10}(x), 
  \label{eq:fsa-s01}
\end{eqnarray}
for the lowest excitation energy $\omega_{10}(N,\delta)$ and transition
amplitude $T_{10}(N,\delta)$.  Numerical results are shown in
figure \ref{fig:e01s01-delta}. Optimal scaling is achieved now with a scaling
variable
\begin{equation}
  \label{eq:x-delta}
  x=N\delta^{\epsilon}, \quad \epsilon=0.72(1),
\end{equation}
which is in good agreement with the value 
\begin{equation}
  \label{eq:2}
  \epsilon=(1+\kappa)^{-1}\simeq0.73(5),
\end{equation}
which follows from the exponent $\kappa=0.37(5)$ in the initial value shown in
figure~\ref{fig:S01-delta0}. This value~(\ref{eq:2}) is compatible with previous
results extracted from a finite-size scaling analysis combined with
renormalization group analysis \cite{Fiel79,SFL86,BB87}.

The exponent $\phi$ in the small $x$-behaviour of the scaling functions
$e_{10}(x)$ and $f_{10}(x)$ is in agreement with Eq.~(\ref{eq:2}):
\begin{equation}
  \label{eq:3}
  \phi=2/\epsilon=2.74(5).
\end{equation}
The coefficients $e_{10}$ and $f_{10}$ in front:
\begin{equation}
  \label{eq:4}
  e_{10}=0.0346(8), \quad f_{10}=-0.0314(8),
\end{equation}
obey the approximate relation~(\ref{eq:e01+f01}).  
It is remarkable to note, that the relation~(\ref{eq:2}) between the exponents
$\epsilon$ and $\kappa$ holds inspite of the fact that the {\it true} large $N$
behaviour $(\kappa=1/2,\; \epsilon=2/3)$ is not yet visible in the finite system
results. Affleck and Bonner \cite{AB90} have discussed the implications of
logarithmic corrections to critical exponents. They have comprised those
corrections into effective scaling dimensions i.e. critical exponents and found
$\epsilon=0.78$ for chain length $N=20$.

In the same way, as we have analysed the large $x$-behaviour of the scaling
function~(\ref{eq:5}) in a staggered field, we have extrapolated the gap
in the presence of a dimerized field:  
\begin{equation}
  \label{eq:delta-gap-x-infinty}
  \Omega_{10}^{\infty} = 6.0(1).
\end{equation}
%
\section{Conclusions and perspectives}
%
In this paper we have developed a general procedure to investigate the behaviour
of antiferromagnetic Heisenberg chains, which are weakly perturbed by an
external field. As examples we considered here a staggered field and a dimerized
field, respectively.

The question of interest is: What happens with the excitation spectrum and
transition amplitudes under the influence of the perturbation? The procedure
used in this paper is based on the observation that excitation energies and
transition amplitudes for the perturbation operator satisfy a system of
differential equations~(\ref{eq:d2En}) and (\ref{eq:dSmn}), which describe the
evolution in the strength $h$ (or $\delta$) of the external field.  The initial
conditions~(\ref{eq:Smn-h0}),~(\ref{eq:Em-E0-h0}) and~(\ref{eq:Sm0-h0}) are
completely fixed by the excitation spectrum and transition amplitudes of the
unperturbed system ($h=0,\delta=0$). We have shown in
section~\ref{sec:scaling-limit} that a scaling ansatz [(\ref{eq:e-s-gap-a}) and
(\ref{eq:e-s-gap-b})] leads to a consistent solution of the evolution equations
in the scaling limit~(\ref{eq:tl}). The exponents $\sigma$ and $\epsilon$ in the
scaling ansatz [(\ref{eq:e-s-gap-a}), (\ref{eq:e-s-gap-b})] and the scaling
variable $Nh^{\epsilon}$ are fixed by the finite-size behaviour of the initial
conditions~(\ref{eq:Smn-h0}),~(\ref{eq:Em-E0-h0}) and~(\ref{eq:Sm0-h0}).

We have also tested numerically the predictions of the scaling ansatz and found
good agreement for the exponents as well as for the scaling functions for small
values of the scaling variable. Therefore, the behaviour of the perturbed system
in the scaling limit~(\ref{eq:tl}) is well understood.

To answer the question whether a gap opens in the presence of an external field
($h$ fixed) in the thermodynamic limit ($N\to\infty$), demands the knowledge of
the scaling functions~(\ref{eq:Omega-x-to-0}) for large values of the scaling
variable. In principle, this asymptotic behaviour follows from a solution of the
differential equations~(\ref{eq:d2Emn-lhs}), (\ref{eq:d2Emn-rhs})
and~(\ref{eq:dlnS_mn-dh}), (\ref{eq:dlnS_mn-dh-lhs}) for the scaling functions
using the initial conditions~(\ref{eq:Omega-x-to-0}) and (\ref{eq:Xi-x-to-0}).
In practice, it might be easier to study this limit numerically coming from
finite system results.  Our numerical data are consistent with a constant
behaviour~(\ref{eq:5}) of the gap scaling function.

The opening of a gap in a uniform field has been recently observed \cite{DHR+97}
in a neutron scattering experiment on copper benzoat $\rm
Cu(C_{6}D_{5}COO)_{2}\cdot3D_{2}O$.  The existence of field dependent and field
independent low energy modes at wave vectors $q=\pi(1-2M)$ and $q=\pi$,
respectively, leads to the expectation, that the compound is adequately
described by the Hamiltonian~(\ref{eq:Hh}) in a uniform external field $B{\bf
  S}_{3}(p=0)$. This system is known to show up zero energy excitations at wave
vectors $q=\pi(1-2M)$ and $q=\pi$ in the longitudinal and transverse structure
factors, respectively \cite{MTBB81,KMS95,FGM+96}.

The exponential fit to the temperature dependence of the specific heat data
revealed, however, that there is a gap which opens with the field strength $B$
as $B^{\epsilon},\; \epsilon\simeq 2/3$. This means of course, that the compound
copper benzoat can not be described by (\ref{eq:Hh}) with $p=0$.  Oshikawa and
Affleck \cite{OA97} argued that the local $g$-tensor for the $\rm Cu$-ions
generates an effective staggered field of strength $h[=h(B)\ll B $]
perpendicular to the uniform field $B$. Therefore, one is lead to investigate
the Hamiltonian:
\begin{equation}
  \label{eq:HhB}
  {\bf H}(h) = {\bf H}+ h{\bf S}_3(0) + B{\bf S}_1(\pi). 
\end{equation}
The authors of reference \cite{OA97} studied the model~(\ref{eq:HhB}) for $B=0$,
which is just the case we investigated in this paper. They came to the same
conclusion concerning the opening of the gap with $h^{\epsilon},\;
\epsilon\simeq 2/3$, as we found here. The question remains, how the exponent
$\epsilon$ changes if the uniform field $B$ is switched on. Moreover, the gap
might evolve in a different fashion for the field dependent and field
independent soft modes $q=\pi(1-2M)$ and $q=\pi$. We will address these
questions in a further publication \cite{FMK98b}.

%
%
%
%

%
%
%

\end{document}